\documentclass[a4paper,11pt]{article}
\usepackage{jheppub} 
\usepackage{lineno}

\arxivnumber{2512.24225} 

\title{\boldmath
Entropic order parameters and
topological holography}

\author[a,b]{Hua-Chen Zhang,}
\author[a]{Germán Sierra,}
\author[c]{and Javier Molina-Vilaplana}
\affiliation[a]{Instituto de F\'isica Te\'orica UAM/CSIC,\\
C/ Nicol\'as Cabrera 13-15, Cantoblanco, 28049 Madrid, Spain}
\affiliation[b]{Asia Pacific Center for Theoretical Physics,\\
77 Cheongam-ro, Nam-gu, Pohang 37673, Korea}
\affiliation[c]{Universidad Politécnica de Cartagena member of European University of Technology EUT+,\\
C/ Dr. Fleming S/N, 30202 Cartagena, Spain}

\emailAdd{huachen.zhang@apctp.org}
\emailAdd{german.sierra@csic.es}
\emailAdd{javi.molina@upct.es}

\abstract{We show that the symmetry topological field theory (SymTFT) construction, also known as the topological holography, provides a natural and intuitive framework for the entropic order parameter characterising phases with (partially) broken symmetries. Various examples of group and non-invertible symmetries are studied. In particular, the origin of the distinguishability of the vacua resulting from spontaneously broken non-invertible symmetries is made manifest with an information-theoretic perspective, where certain operators in the SymTFT are excluded from observation.}


\begin{document}
\maketitle
\flushbottom

\section{Introduction and summary}

Understanding the phases of quantum field theories and condensed matter with strong couplings is a long-standing core problem in physics, prompting the pursuit of principles that transcend the perturbative realm. Symmetry is one such principle; indeed, it has served as a fundamental organising concept since Landau's celebrated theory, where the phases are classified by how their symmetries manifest at low energy. The introduction of `generalised' symmetries, which include, most notably, higher-form symmetries~\cite{gaiotto2015} and non-invertible (or fusion category) symmetries~\cite{bhardwaj2018}, has substantially enriched the zoo of phases arising from symmetry breaking and symmetry protection.

In the original Landau paradigm, the symmetry breaking is generally signalled by the non-vanishing expectation value of a certain (local) operator, which is termed the order parameter. One drawback of this `traditional' notion of order parameter is that it is {\it a priori} not clear {\it which} operator is pertinent to a certain situation, which is furthermore not unique; in particular, the operator order parameter is notoriously difficult to identify for systems with strong couplings. To this end, new kinds of characteristic quantities based on information theory, termed entropic order parameters~\cite{casini2020,magan2021,casini2021,magan2021a,molina2024,molina2025} or entanglement asymmetry~\cite{marvian2014,capizzi2023,fossati2024,kusuki2025,fossati2025,ares2023,benini2025}, were advocated. Mathematically, this quantity is a relative entropy quantifying the distinguishability of two states~\cite{vaccaro2008,gour2009}.

The problems regarding relative entropies in quantum field theory are conveniently approached using the Haag-Kastler algebraic formalism~\cite{haag1964,haag1996}, where one considers operator algebras generated by quantum fields smeared with test functions supported in certain regions of the spacetime. The theory is specified by a causal net of algebras, which assigns an operator algebra to each open spacetime region. Ref.~\cite{casini2020} presented a detailed discussion of situations in which the theory contains superselection sectors. It was demonstrated that different subalgebras can be assigned to topologically non-trivial regions, where non-locally generated operators are included or excluded, respectively. (See also ref.~\cite{shao2025}.) The difference between these choices of operator algebras, in turn, is quantified by the difference in mutual information, which is well-defined in the continuum and can be expressed as a relative entropy that discerns the existence of the superselection sectors. The latter form a tensor category, which is equivalent to the representation category $\mathrm{Rep}(G)$ of a compact group $G$. $\mathrm{Rep}(G)$ can be regarded as the~\emph{dual} (non-invertible) symmetry obtained by~\emph{orbifolding} (or equivalently,~\emph{gauging}) the original symmetry $G$~\cite{bhardwaj2018}. As we shall see, this situation is closely related to that of spontaneous symmetry breaking (SSB), where non-trivial relative entropies can be defined already for a topologically trivial region (i.e. that with the topology of a single disk). The SSB of generalised symmetries gives rise to interesting new features. In ref.~\cite{benini2025}, it was shown that the relative entropies for distinct vacua arising from the SSB of non-invertible symmetries can be different; physically, this indicates that these vacua are~\emph{distinguishable}.

In this work, we study relative entropies as order parameters for phases of spontaneously broken invertible and non-invertible symmetries with the aid of an additional instrument known as topological holography or symmetry topological field theory (SymTFT) \cite{ji2020,kong2020,gaiotto2021,burbano2022,chatterjee2023,freed2024,kaidi2023a,kaidi2023b,bhardwaj2025}. The latter encodes the (generalised) symmetry of a quantum field theory (QFT) holographically by using a topological field theory (or topological order) living in a spacetime that is one dimension higher than the original QFT. The power of this approach lies in its neat separation of the properties that solely derive from the symmetry (i.e. the `kinematics') from the dynamics. We will see that the definition and computation of entropic order parameters for SSB phases naturally fit into the SymTFT construction, which keeps track of the non-locally generated operators (the `intertwiners' introduced below) responsible for the relative entropies in an intuitive way. For the SSB of ordinary (invertible) symmetries, these intertwiners form a representation of the symmetry group; the vacua, on the other hand, correspond to a special basis, namely that of idempotents, in the representation space. Minimal examples for Abelian groups, $\mathbb{Z}_{2}$ with or without 't Hooft anomaly, and $\mathbb{Z}_{2} \times \mathbb{Z}_{2}$, are examined to illustrate the approach. The symmetry $\mathbb{Z}_{2} \times \mathbb{Z}_{2}$ exhibits a non-trivial symmetry-protected topological (SPT) phase~\cite{gu2009,pollmann2010,pollmann2012,chen2013}; as we will see, while there is no non-vanishing entropic order parameter, the non-trivial SPT vacuum corresponds to a `twisted sector' counterpart of the intertwiner. For non-Abelian group symmetries, we consider the example $S_{3}$. Another nice feature of this approach is that mutually Morita dual symmetry categories share the same SymTFT; the most basic example is given by the pair $(\mathrm{Vec}_{G},\mathrm{Rep}(G))$.\footnote{The category of $G$-graded vector spaces, $\mathrm{Vec}_{G}$, is the symmetry category of the group symmetry $G$.} $\mathrm{Rep}(G)$ for a non-Abelian group $G$ is called a~\emph{group-theoretical} non-invertible symmetry. In relation to the example of $S_{3}$, we will consider the SSB phases of $\mathrm{Rep}(S_{3})$. Finally, going beyond group-theoretical symmetries, we will study the case of the non-invertible symmetry described by the Ising fusion category. In the situation of non-invertible symmetries, the notion of a representation is replaced by that of a module category. In both the invertible and the non-invertible cases, the vacua are expressed as linear combinations of the intertwiners; we emphasise its close analogy with the expression of Cardy states in terms of Ishibashi states in rational conformal field theories~\cite{kishimoto2004}, where the idempotency is readily checked using the Verlinde formula~\cite{verlinde1988}. This is not surprising, as boundary states and SSB vacua both admit the mathematical structure of module categories. The basis of idempotents eases the computation of the relative entropy, which quantifies the information loss when the non-trivial intertwiners are excluded from observation in each vacuum. The values of this quantity are shown to be directly related to the quantum dimensions of simple objects in the fusion category associated with the broken symmetry. Remarkably, they can be different for non-invertible symmetries, accounting for the distinguishability of the corresponding vacua. Gapped (1+1)d phases with non-invertible symmetries were studied using SymTFT techniques in refs.~\cite{bhardwaj2025a,bhardwaj2024}, where the distinguishability of the vacua was observed from a different perspective; we shall come back to this point later.

The relevant notions, including symmetry-breaking phases, SymTFT, and entropic order parameter, are reviewed in more detail in section~\ref{sec:review}. The examples of group symmetries and non-invertible symmetries are studied in section~\ref{sec:invertible} and section~\ref{sec:noninvertible}, respectively. While our focus in the present work is on the gapped phases in (1+1) dimensions, we note that the SymTFT construction facilitates the generalisation to gapless phases~\cite{vanhove2018,vanhove2022,vanhove2022a,huang2025,bhardwaj2025c,bhardwaj2025d,bhardwaj2024a,bottini2025,hung2025,luo2025} and higher-dimensional systems~\cite{bhardwaj2025e,antinucci2025,bhardwaj2025f,inamura2025}; in particular, the SSB of 1-form symmetries is pertinent to (2+1)d topological orders~\cite{mcgreevy2023}. The investigation of entropic order parameters for these situations will be reported in the future.

\section{Symmetry-breaking phases, SymTFT, and entropic order parameter}
\label{sec:review}

We consider a 2d [or, equivalently, (1+1)d] QFT with a global symmetry, which can be non-invertible and described by a fusion category $\mathcal{C}$. The latter acts on $\mathcal{F}$, the algebra generated by the full operator content of the theory. Mathematically, each gapped phase for such a QFT corresponds to a way of~\emph{gauging} the global symmetry, which, in turn, is specified by an indecomposable~\emph{module category} over $\mathcal{C}$~\cite{bhardwaj2018,choi2024}. In particular, the theory that has operator algebra $\mathcal{O}$, which is the subalgebra of $\mathcal{F}$ generated by the operators invariant (i.e. `uncharged') under $\mathcal{C}$, is obtained by the gauging with respect to a canonical choice of the module category with only one simple object.\footnote{In the literature, $\mathcal{F}$ and $\mathcal{O}$ are sometimes termed the~\emph{field algebra} and the~\emph{observable algebra}, respectively.} As a familiar example, for the theory $\mathcal{F}$ with finite group symmetry $\mathrm{Vec}_{G}$, the $G$-\emph{orbifolded} theory $\mathcal{O}$ results from gauging $\mathrm{Vec}_{G}$ with respect to the module category $\mathrm{Vec}$ (the category of vector spaces) and has the `dual' non-invertible symmetry $\mathrm{Rep}(G)$; the latter describes the superselection sectors of the orbifolded theory. If, instead of the whole symmetry group $G$, only a (non-anomalous) subgroup $H<G$ is gauged or orbifolded, the resulting operator algebra (which is larger than $\mathcal{O}$) is only invariant under the subgroup $H$; this is the case of a partially symmetry-breaking phase, where $H$ is the unbroken subgroup. We will see that this story carries over to general non-invertible cases.

More specifically, we focus on~\emph{gapped} phases, for which the low-energy effective theories are~\emph{topological} quantum field theories (TQFTs). Consider the (1+1)d theory on an infinitely long cylinder, where the time direction is along the axis. The Hilbert space associated with the spatial circle $S^{1}$ is finite dimensional; the~\emph{vacua}
\begin{equation}
    \{ v_{1}, v_{2}, \ldots, v_{N} \}
\end{equation}
form a basis of this Hilbert space and are indexed by the simple objects of a module category over $\mathcal{C}$. Under the radial map from the cylinder to the plane, each vacuum is mapped to an operator; the operator product expansion (OPE) takes the simple form
\begin{equation}
    v_{k}v_{k^{\prime}} = \delta_{kk^{\prime}}v_{k}, \quad k, k^{\prime} = 1, \ldots, N.
\end{equation}
Namely, the vacua are orthogonal idempotents. Two familiar extreme cases are i) an SPT phase, where $N=1$\footnote{A module category with only one simple object is equivalent to a~\emph{fibre functor}: $\mathcal{C} \rightarrow \mathrm{Vec}$~\cite{etingof2015}.}, and ii) the gapped phase with a symmetry group $G$ completely broken, for which the module category reduces to the regular representation, where the basis vectors are in one-to-one correspondence with the elements of $G$ itself, and $N = |G|$.

\begin{figure}[htbp]
\centering
\includegraphics[width=\textwidth]{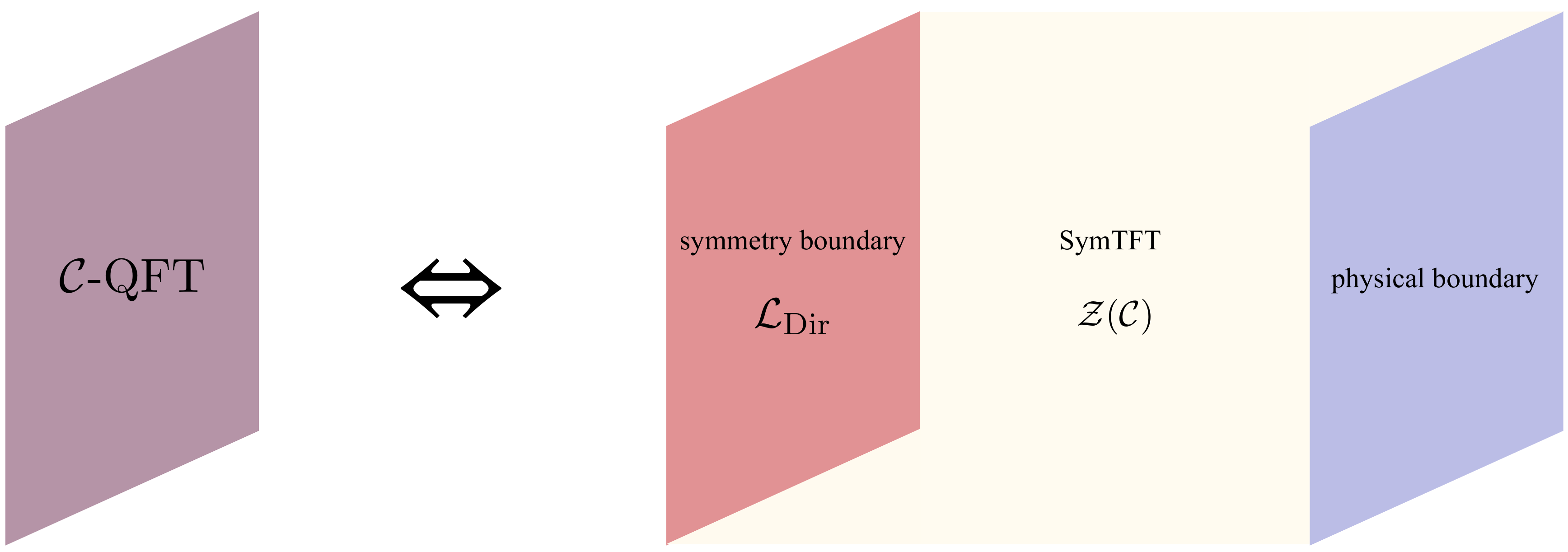}
\caption{The 2d QFT with symmetry category $\mathcal{C}$ is represented as a `sandwich' consisting of i) the 3d SymTFT in which the topological lines form $\mathcal{Z}(\mathcal{C})$, ii) the symmetry boundary where the SymTFT has Dirichlet boundary condition, and iii) the physical boundary which is not necessarily topological. \label{fig:1}}
\end{figure}

Let us see how these symmetry-breaking phases emerge in the SymTFT construction. The latter represents a 2d QFT with symmetry category $\mathcal{C}$ as a `sandwich' consisting of a 3d topological field theory (the~`\emph{SymTFT}') and two boundaries thereof, see figure~\ref{fig:1}; the original 2d theory is recovered by compactifying the sandwich and bringing the two boundaries together. The topological line operators (i.e. `anyons') in the SymTFT form the modular category $\mathcal{Z}(\mathcal{C})$, which is the Drinfeld centre of $\mathcal{C}$~\cite{turaev1992,levin2005}. One of the boundaries of the sandwich, the~\emph{symmetry boundary}, is a particular topological boundary condition called Dirichlet. A topological boundary of the SymTFT is again specified by a module category over $\mathcal{C}$; the Dirichlet boundary condition corresponds to the regular module category. The topological lines living on the Dirichlet boundary form the fusion category $\mathcal{C}$, hence the name `symmetry boundary'. The other boundary, which is not necessarily topological, encodes the dynamics of the theory and is called the~\emph{physical boundary}. For TQFTs pertinent to gapped phases, the physical boundary is chosen to be topological as well. We have therefore recovered the fact that the $\mathcal{C}$-gapped phases are classified by module categories over $\mathcal{C}$, which correspond to distinct physical boundaries of the SymTFT. The SymTFT construction offers extra convenience, largely because it intuitively provides an alternative characterisation of the topological boundaries, which is equivalent to that in terms of module categories. This is done by specifying the topological lines in the SymTFT that can condense (namely, end topologically) on the boundary, which are encoded in a~\emph{Lagrangian algebra}
\begin{equation}
\mathcal{L} = \bigoplus_{\mathfrak{a}\in \mathcal{Z}(\mathcal{C})} n_{\mathfrak{a}} \mathfrak{a}
\end{equation}
in the Drinfeld centre, where $\mathfrak{a}$ denotes the simple topological lines, and $n_{\mathfrak{a}}$ are non-negative integers subject to certain conditions~\cite{lan2015,davydov2017,cong2016}.\footnote{In the terminology of anyon condensation~\cite{bais2009,kong2014,neupert2016}, a Lagrangian algebra is a condensable algebra containing certain (bosonic) anyons, of which the condensation fully confines the topological order. Non-Lagrangian condensable algebras are associated with topological~\emph{interfaces} instead of boundaries.} In particular, the symmetry boundary is associated with $\mathcal{L}_{\mathrm{Dir}}$, the Lagrangian algebra for the Dirichlet boundary condition. Gapped SSB phases are cases where non-trivial topological lines in the SymTFT can condense on both the symmetry and physical boundaries; after compactifying the SymTFT sandwich, the resulting local operators are mapped to the symmetry-broken vacua under the state-operator correspondence. We will see concrete examples of these in sections~\ref{sec:invertible} and~\ref{sec:noninvertible}.

\begin{figure}[htbp]
\centering
\includegraphics[width=\textwidth]{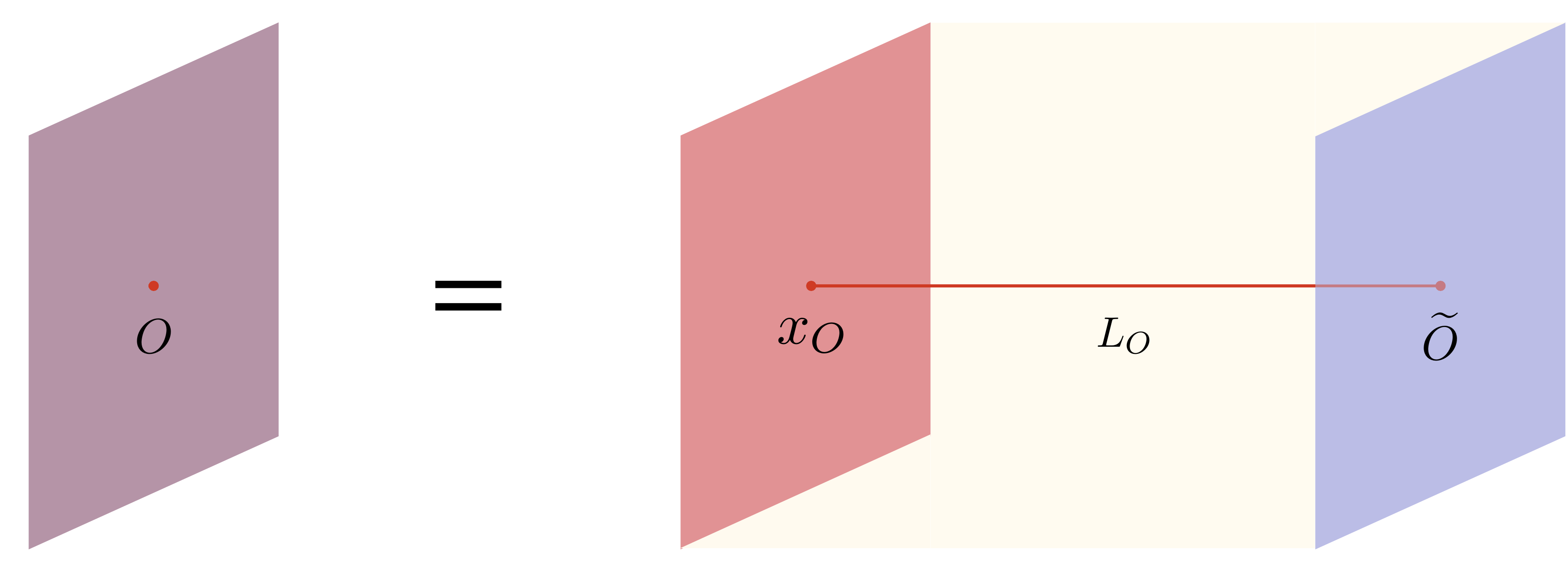}
\caption{In the SymTFT construction, a local operator $O \in \mathcal{F}$ is represented as a triple $(x_{O}, L_{O}, \widetilde{O})$, where $L_{O}$ is a topological line in the SymTFT, $x_{O}$ is the topological junction of $L_{O}$ on the symmetry boundary, and $\widetilde{O}$ is its (generally non-topological) junction on the physical boundary. \label{fig:2}}
\end{figure}

\begin{figure}[htbp]
\centering
\includegraphics[width=\textwidth]{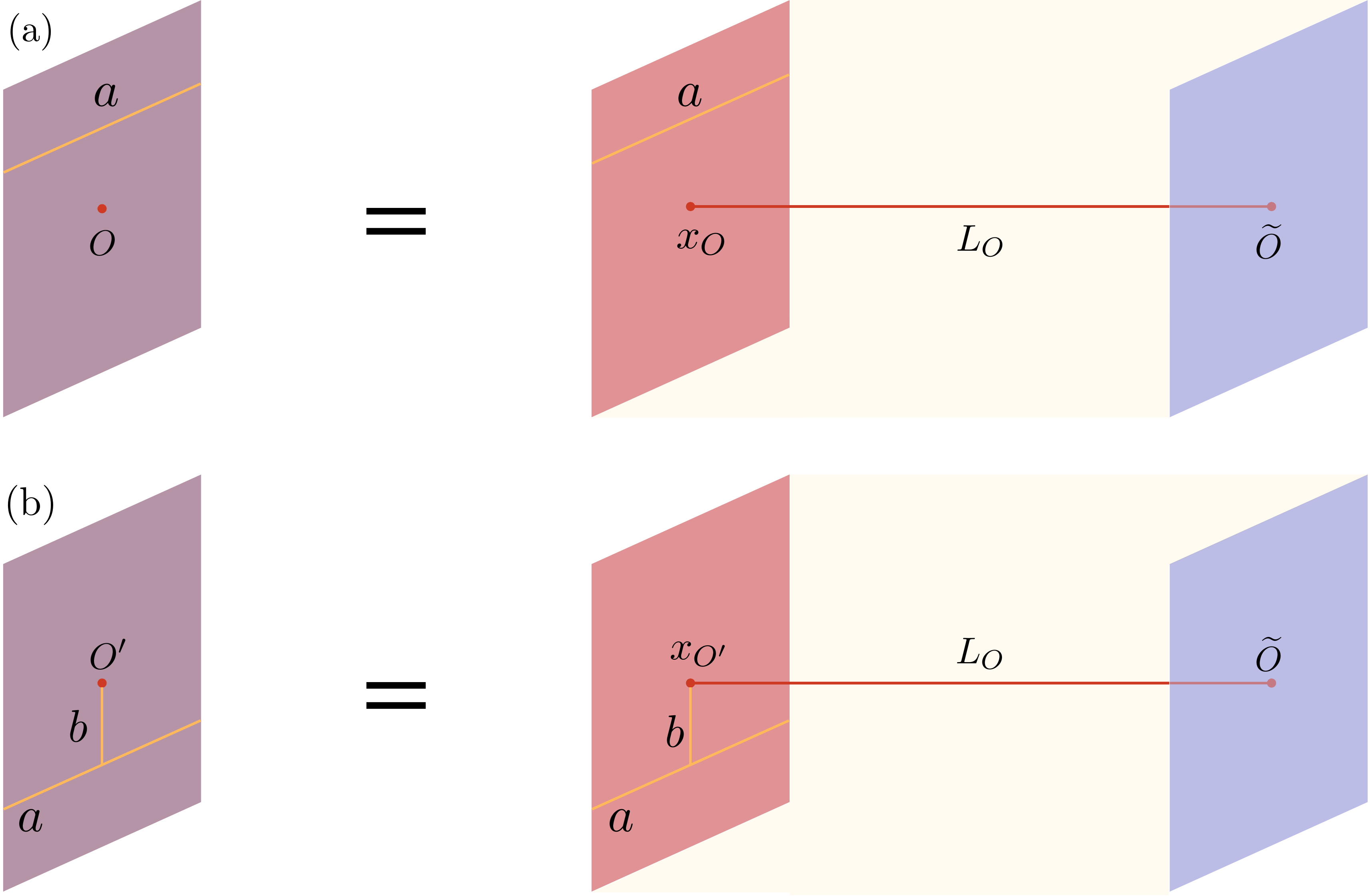}
\caption{The action of $a$ in the symmetry category $\mathcal{C}$ on a (genuinely local) operator $O \in \mathcal{F}$ amounts to pushing the corresponding topological defect line on the symmetry boundary through the junction $x_{O}$. This action can leave behind another defect line $b$ attaching the junction to the line $a$. Note that actions of the global symmetry are performed solely on the symmetry boundary and cannot change the bulk topological line $L_{O}$. The latter labels the multiplet in which the operator $O$ transforms. In particular, if $L_{O}$ is the trivial line, the corresponding operator $O$ is invariant under the symmetry and belongs to $\mathcal{O}$. \label{fig:3}}
\end{figure}

In the SymTFT construction, a local operator $O \in \mathcal{F}$ is represented as a triple $(x_{O}, L_{O}, \widetilde{O})$, see figure~\ref{fig:2}. Here, $L_{O}$ is a linear combination of simple topological lines $\mathfrak{a}$ in the SymTFT with non-negative coefficients. $L_{O}$ can end topologically on the symmetry boundary, and $x_{O}$ is a topological operator at the junction. $\widetilde{O}$, on the other hand, is the junction operator of $L_{O}$ on the physical boundary; as the latter does not have to be a topological boundary condition, $\widetilde{O}$ is generally non-topological. The action of $a$ in the symmetry category $\mathcal{C}$ on $O$  amounts to pushing the corresponding topological defect line on the symmetry boundary through the junction $x_{O}$. For a non-invertible symmetry, the above action can leave behind a defect line attaching the junction to the line $a$. The situation is depicted in figure~\ref{fig:3}. As it is clear that actions of the global symmetry cannot change the bulk topological line $L_{O}$, we reach the conclusion that $L_{O}$ labels the multiplet in which the operator $O$ transforms. In particular, an invariant operator $O \in \mathcal{O}$ is one for which $L_{O}$ is the trivial line in the SymTFT.

A remark is in order. The operator $O$ is a genuinely local operator, which refers to one that is not attached to a non-trivial topological defect line. On the contrary, an operator residing at the end of a semi-infinite non-trivial line is said to be in a twisted sector. This is easily understood using the state-operator correspondence, where the operator is mapped to a state living in the Hilbert space twisted by the defect line. We take the point of view that the algebra $\mathcal{F}$ does not include these twisted-sector operators. The latter, however, play an important role in the characterisation of SPT phases. We have seen that the operators in a multiplet for a non-invertible symmetry can live in different twisted sectors.

\begin{figure}[htbp]
\centering
\includegraphics[width=\textwidth]{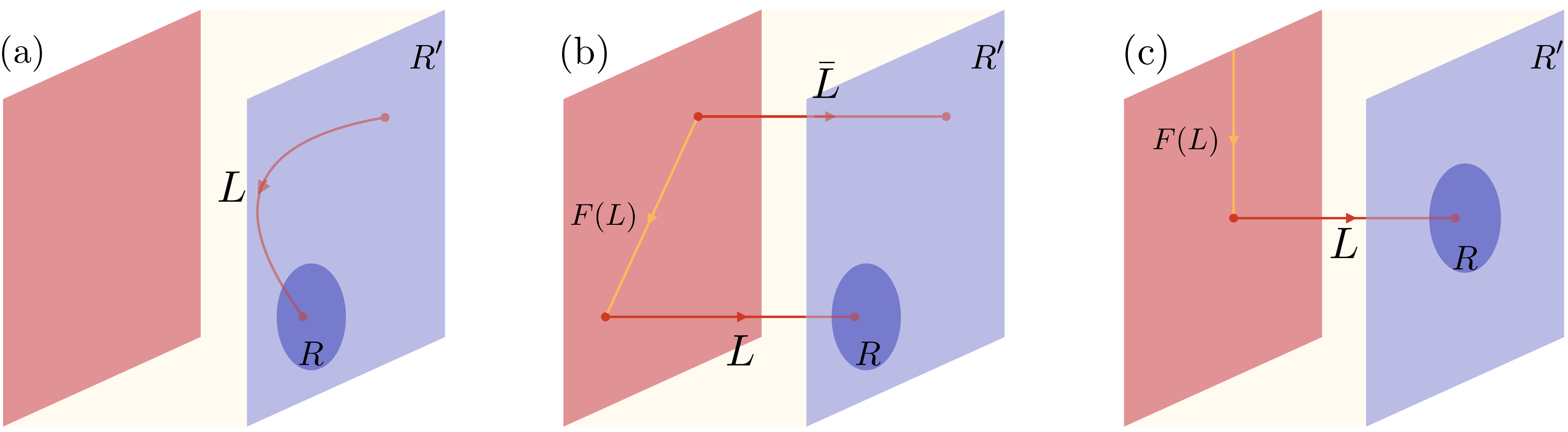}
\caption{SymTFT representation of the intertwiner $\mathcal{I}_{RR^{\prime}}$ supported on the region $R$ and its complement $R^{\prime}$. It is evident from (a) that $\mathcal{I}_{RR^{\prime}}$ is invariant under the global symmetry and belongs to $\mathcal{O}$. In (b), the bulk topological line $L$ is merged into the symmetry boundary using the bulk-to-boundary map $F$; $\bar{L}$ is the orientation reversal of $L$. In (c), the line $\bar{L}$ and the associated junction are sent to infinity; the operator resulting in this limit is denoted as $\mathcal{I}_{R}$, which lives in the $F(L)$-twisted sector. We will mainly focus on the case where $F(L)$ is the identity line, for which $\mathcal{I}_{R}$ belongs to $\mathcal{F}(R)$ but not to $\mathcal{O}(R)$. \label{fig:4}}
\end{figure}

One type of operators that will be crucial in our discussions is known as the~\emph{intertwiners}, which are defined as follows.\footnote{In an important alternative perspective, one focuses on the localised Doplicher-Haag-Roberts (DHR) endomorphisms of the algebra $\mathcal{O}$~\cite{doplicher1969a,doplicher1969b,doplicher1971,doplicher1974}; starting with the vacuum representation of $\mathcal{O}$, the other representations can be constructed using these endomorphisms. The intertwiners map between the DHR endomorphisms. In the text, the terminology `intertwiner' is used in a more general sense. For group symmetries, the intertwiners reduce to the more familiar ones labelled by the representations; see section~\ref{sec:invertible}.} Suppose $R$, which has the topology of a disk, is an open causal region in the 2d spacetime; $R^{\prime}$ is its causal complement. $\mathcal{F}(R)$ [resp. $\mathcal{O}(R)$] denotes the subalgebra of $\mathcal{F}$ (resp. $\mathcal{O}$) formed by the operators locally generated in $R$. An intertwiner $\mathcal{I}_{RR^{\prime}}$, associated with a bulk topological line $L$, is an operator in $\mathcal{O}$ as depicted in figure~\ref{fig:4}(a). We can also merge the bulk line $L$ into the symmetry boundary using the bulk-to-boundary map\footnote{The bulk-to-boundary map is also known as the~\emph{forgetful functor}, which simply describes the result one gets when a topological line in the SymTFT is pushed into the Dirichlet boundary. In general, a simple topological line in the bulk SymTFT can be mapped to a non-simple one on the symmetry boundary.} $F: \mathcal{Z}(\mathcal{C}) \rightarrow \mathcal{C}$ as shown in figure~\ref{fig:4}(b), where $\bar{L}$ is the orientation reversal of $L$. Importantly, $\mathcal{I}_{RR^{\prime}}$ is not locally generated in $R$ and does not belong to $\mathcal{O}(R)$. If we focus on the subalgebra assigned to the region $R$, we can imagine sending the topological line $\bar{L}$ and the associated junction to infinity\footnote{This is similar to creating a pair of particles with opposite charges from the vacuum and transporting one of them to infinity. We have assumed implicitly that the charges are~\emph{transportable}.}; in this limit, the resulting operator obtained by compactifying the SymTFT sandwich is denoted as $\mathcal{I}_{R}$, which lives in the $F(L)$-twisted sector; see figure~\ref{fig:4}(c). By slightly abusing the terminology, we also refer to these operators as `intertwiners'. When $F(L)$ is the identity line, the `untwisted' $\mathcal{I}_{R}$ belongs to $\mathcal{F}(R)$. We will mainly focus on these untwisted intertwiners for characterising the SSB phases, although, as alluded to in the remark above, the twisted-sector counterparts are pertinent to the SPT phases. Let us note that the intertwiners we study here are essentially the same as what are called~\emph{patch operators} in refs.~\cite{chatterjee2023,lan2024,inamura2023,schafer2025,jia2025}; in fact, it was already conjectured~\cite{inamura2023} that certain patch operators would serve as order parameters for gapped phases. Our emphasis, however, is on the non-locally generated nature of these operators and the consequences they have for the relative entropies, which we will elaborate on next.

\subsection{Relative entropy as order parameter}
\label{subsec:entropic-order-parameter}

The relative entropy~\cite{umegaki1962} establishes a distance between two quantum states by quantifying their distinguishability. Let $\varphi_{1}$ and $\varphi_{2}$ be two states (i.e. positive linear functionals) on an algebra $\mathcal{M}$ of $N \times N$ matrices, the relative entropy between them is defined in terms of the corresponding density matrices as
\begin{equation}
\label{eq:relative-entropy-def}
    S(\varphi_{1}|\varphi_{2})~{:=}~\mathrm{Tr}\left[\rho_{\varphi_{1}}\left(\log{\rho_{\varphi_{1}}} - \log{\rho_{\varphi_{2}}}\right)\right].
\end{equation}
Note that the von Neumann entropy $S(\varphi) {:=} -\mathrm{Tr}(\rho_{\varphi}\log{\rho_{\varphi}})$ of a state $\varphi$ is directly related to its relative entropy $S(\varphi|\tau)$ with the maximally mixed state $\rho_{\tau} = N^{-1}\mathbb{I}$,
\begin{equation}
    S(\varphi) = \log{N} - S(\varphi|\tau).
\end{equation}
We observe that $\tau$ can be expressed in terms of a linear map $E: \mathcal{M} \rightarrow \mathfrak{I}$ with $E(m) = N^{-1} \mathrm{Tr}(m) \mathbb{I}$ as $\tau = \varphi \circ E$ for any state $\varphi$, where $\mathfrak{I}$ is the subalgebra of $\mathcal{M}$ generated by $\mathbb{I}$. This map $E$ is an example of a~\emph{conditional expectation}. More generally, for any subalgebra $\mathcal{N}$ of $\mathcal{M}$, a conditional expectation is a positive linear map $E: \mathcal{M} \rightarrow \mathcal{N}$ satisfying the conditions
\begin{equation}
    E(\mathbb{I}) = \mathbb{I},\quad E(n_{1}mn_{2}) = n_{1}E(m)n_{2},~\forall~ m \in \mathcal{M}\text{~and~}n_{1}, n_{2} \in \mathcal{N}.
\end{equation}
In particular, the second equality above is known as the bimodule property. If the conditional expectation preserves the trace, i.e. $\mathrm{Tr}(E(m)) = \mathrm{Tr}(m)~\forall m \in \mathcal{M}$, it is readily proved that~\cite{casini2020,casini2021}
\begin{equation}
\label{eq:relative-entropy-trace-preserving}
    S(\varphi|\varphi \circ E) = S(\varphi \circ E) - S(\varphi).
\end{equation}
Namely, in this case, the relative entropy $S(\varphi|\varphi \circ E)$ is expressed as the difference between the von Neumann entropies of the two states.

Now, with the SymTFT picture in mind, imagine bringing a genuinely local operator in $\mathcal{O}(R)$ close to the junction associated with $\mathcal{I}_{R}$; the OPE yields the composition, which is clearly also attached to the same bulk line $L$. Thus, this OPE produces a module denoted as $[\mathcal{I}_{R}]$ over such local operators; the different modules close a fusion algebra. For notational simplicity, we conflate the intertwiner $\mathcal{I}_{R}$ and its associated class $[\mathcal{I}_{R}]$. If we define $E: \mathcal{F}(R) \rightarrow \mathcal{O}(R)$ as the positive linear map that removes all non-invariant parts of a local operator under the symmetry $\mathcal{C}$ (i.e. those that are attached to a non-trivial bulk line in the SymTFT representation), this map satisfies the condition
\begin{equation}  E(O^{\mathrm{inv}}_{1}~O~O^{\mathrm{inv}}_{2}) = O^{\mathrm{inv}}_{1} E(O) O^{\mathrm{inv}}_{2},~\forall~ O \in \mathcal{F}(R)\text{~and~}O^{\mathrm{inv}}_{1}, O^{\mathrm{inv}}_{2} \in \mathcal{O}(R),
\end{equation}
which means that the operation of taking the OPE with invariant operators commutes with that of acting by $E$ and is precisely the bimodule property. This becomes manifest in the SymTFT representation, as the OPE takes place on the physical boundary and cannot alter the bulk line associated with the operator $O$. In fact, such a map $E$ is a conditional expectation. Starting from a state on $\mathcal{F}(R)$, one can form another `symmetrised' one using the conditional expectation. In our applications, the state is taken to be one of the vacua $v_{k}$, which is expressed as a linear combination of the intertwiners; schematically, we write (the subscript $R$ of the intertwiners indicating the region is omitted when it does not cause ambiguity)
\begin{equation}
\label{eq:vacua-expression-schematic}
    v_{k} = \sum_{x} C_{k,x} \mathcal{I}^{(x)},
\end{equation}
where $x$ is a collective label for the intertwiners and their components (see below). As the vacua form an orthonormal basis, the density matrix of $v_{k}$ in this basis is diagonal with diagonal elements $\langle v_{k} \rangle_{k^{\prime}} = \delta_{kk^{\prime}}$, i.e., they are pure states on $\mathcal{F}$ with deterministic probability distribution of outcomes of measuring. The conditional expectation `kills' the non-trivial intertwiners: $v_{k} \mapsto C_{k,0} \mathcal{I}^{(0)}$, where $\mathcal{I}^{(0)}$ denotes the trivial intertwiner; in other words, 
\begin{equation}
    \rho_{v_{k} \circ E} = C_{k,0} \mathbb{I}.
\end{equation}

 First, we notice that $E$ leaves the (only) symmetric vacuum intact, and the resulting relative entropy vanishes trivially. Hence, the relative entropies $S(v_{k} | v_{k} \circ E)$ serve as the~\emph{entropic order parameter} signalling spontaneous symmetry breaking. The second simplest case is when $E$ is trace-preserving, for which it is clear that $C_{k,0} = N^{-1}$ ($N$ is the number of vacua); from~\eqref{eq:relative-entropy-trace-preserving} it follows that $S(v_{k} | v_{k} \circ E) = \log{N}$, which is the same for each vacuum. This is the case of an invertible symmetry, where one projects to the component of an operator that is invariant under the symmetry group. The breaking of non-invertible symmetries, on the other hand, is associated with conditional expectations that are~\emph{not} trace-preserving. As we shall see, the relative entropies for different vacua can then differ, rendering them distinguishable.

Before leaving this section, let us remark on the connection with the discussions in ref.~\cite{bhardwaj2025a,bhardwaj2024}, regarding what is called \emph{Categorical Landau Paradigm}, which extends the Landau paradigm of symmetry breaking to the case of non-invertible symmetries. There, the authors noted that in (1+1)d gapped phases, the vacua arising from the spontaneous breaking of non-invertible symmetries may be physically distinguishable, contrary to the case of invertible symmetries. This depends on the properties of interfaces (which are line defects) between different vacua and the linking action of such interfaces on the vacua, which yields their~\emph{relative Euler terms}; non-vanishing relative Euler terms indicate the distinguishability of the vacua.

Here, we argue that the relative Euler term between two vacua is directly related to the eigenvalues of the relative modular operator for the corresponding symmetrised states. Thereby, we provide an algebraic characterization of the distinguishability of vacua in the symmetry breaking of non-invertible symmetries. To see this, we note that for each state $v_{k} \circ E$ on $\mathcal{F}$, one can perform a Gelfand-Naimark-Segal construction~\cite{gelfand1943,segal1947}, which induces a representation of $\mathcal{F}$ with a unit cyclic vector $| k \rangle$; there are linear operators $\mathfrak{S}_{k^{\prime}k}$ satisfying
\begin{equation}
    \mathfrak{S}_{k^{\prime}k} | k \rangle = \mathrm{e}^{-(\lambda_{k^{\prime}} - \lambda_{k})} | k^{\prime} \rangle,
\end{equation}
where the relative Euler term between the pair of vacua is simply $\lambda_{k^{\prime},k} \equiv \lambda_{k^{\prime}} - \lambda_{k}$ appearing in the coefficient. This operator admits a polar decomposition
\begin{equation}
\mathfrak{S}_{k^{\prime}k} = J_{k^{\prime}k} \Delta_{k^{\prime}k}^{1/2},
\end{equation}
where $J_{k^{\prime}k}$ is the modular conjugation between $| k \rangle$ and $| k^{\prime} \rangle$ and
\begin{equation}
    \Delta_{k^{\prime}k} = \rho_{v_{k^{\prime}} \circ E}~(\rho_{v_{k} \circ E})^{-1} = \frac{C_{k^{\prime},0}}{C_{k,0}} \mathbb{I}
\end{equation}
is the relative modular operator. We note that the relative entropy can be defined in terms of the  modular operators through the Araki formula \cite{Araki:1976zv}. This constitutes  an operator-algebraic definition of relative entropy for normal states on von Neumann algebras. It defines a robust, non-negative, and monotonic measure of distinguishability between two states. In our setting, $\rho_{v_{k} \circ E}$ and $\rho_{v_{k^{\prime}} \circ E}$ are simultaneously diagonalised in the basis spanned by the vacua, hence one has 
\begin{equation}
    \lambda_{k^{\prime},k} = \frac{1}{2} \log\, \left(\frac{C_{k,0}}{C_{k^{\prime},0}}\right) \quad \Rightarrow \quad \mathfrak{S}_{k^{\prime}k} | k \rangle = \mathrm{e}^{-\lambda_{k^{\prime},k}} | k^{\prime} \rangle = \sqrt{\frac{C_{k^{\prime},0}}{C_{k,0}}}\, | k^{\prime} \rangle\, ;
\end{equation}
again, one finds that the relative Euler term vanishes if the conditional expectation is trace-preserving.

\section{Group symmetries}
\label{sec:invertible}

We first focus on the simpler and more familiar examples of invertible (or `ordinary') symmetries described by a finite group $G$. In this case, the Drinfeld centre reduces to the quantum double of $\mathrm{Vec}_{G}$. The corresponding SymTFT is known as the 3d Dijkgraaf-Witten (DW) gauge theory~\cite{dijkgraaf1990}, where a simple topological line is labelled by a pair
\begin{equation}
    ([g],r),
\end{equation}
where $[g]$ is a conjugacy class in $G$, and $r$ is an irreducible representation (irrep) of the centraliser $\mathsf{C}_{G}(g)$ of a representative element $g$ in this class.\footnote{For simplicity, we assume here that the symmetry $G$ is non-anomalous. The SymTFT for $\mathrm{Vec}_{G}^{\omega}$ with an 't Hooft anomaly $\omega \in \mathrm{H}^{3}(G,U(1))$ is the~\emph{twisted} DW theory, where the irreps of the centralisers have twists that are induced by $\omega$. We will see a simple example of anomalous $\mathbb{Z}_{2}$ symmetry below.} The symmetry boundary corresponds to the Dirichlet boundary condition of this DW theory, which is described by the Lagrangian algebra
\begin{equation}
    \mathcal{L}_{\mathrm{Dir}} = \bigoplus_{r} ([e],r).
\end{equation}
Namely, the topological lines of the form $([e],r)$ (i.e. the Wilson lines) can condense on the symmetry boundary, where $e$ is the unit element and $r$ is an irrep of $G$. Therefore, an untwisted-sector intertwiner $\mathcal{I}^{r}$ is associated with a bulk topological line $([e],r)$ in the SymTFT representation and is labelled by an irrep $r$.\footnote{As discussed in section~\ref{sec:review}, $\mathcal{I}^{r}$ can also be regarded as a module over the operator algebra $\mathcal{O}(R)$. This double role played by the label $r$ is known as the Schur-Weyl duality~\cite{choi2024}.} Note that these intertwiners are allowed by the symmetry of the theory (i.e. the `\emph{kinematics}'); whether these operators actually exist in the theory is, of course, dictated by the~\emph{dynamics} encoded on the physical boundary. The latter is precisely the information of symmetry breaking in our context. As alluded to in section~\ref{sec:review}, the choice of a topological (gapped) physical boundary condition is equivalent to that of a module category over the symmetry category. This is also reflected in the bulk lines that can end topologically on both the symmetry and the physical boundaries, which correspond precisely to the (untwisted) intertwiners that exist in the theory. The components of these intertwiners correspond to the simple objects of the module category -- note that a simple bulk line can end on a topological boundary in different ways, leading to components of the intertwiner. In the case of a group symmetry, these reduce to the basis vectors in an irrep, which, in general, appear multiple times. Specifically, the physical boundary is the Neumann boundary condition for the gauge fields corresponding to a subgroup $H$ of $G$, which is the symmetry left unbroken\footnote{More accurately, in addition to the subgroup $H$ of $G$, the physical boundary (or module category) is associated with an element in $\mathrm{H}^{2}(H,U(1))$ known as the~\emph{discrete torsion}. For $H<G$, a non-trivial discrete torsion implies spontaneous symmetry breaking mixed with symmetry protection. This situation will not be considered in the following; nonetheless, we will look at the example $H = G = \mathbb{Z}_{2} \times \mathbb{Z}_{2}$, namely the non-trivial $\mathbb{Z}_{2} \times \mathbb{Z}_{2}$ SPT phase.}; this amounts to the regular representation of the broken symmetry, where an irrep $r$ with dimension $d_{r}$ appears $d_{r}$ times. (It is worth noting that irreps for the full group $G$ generally become reducible and decompose into different $r$s.) The $i$-th basis vector in the $j$-th copy of $r$ corresponds to the component $\mathcal{I}^{r}_{ij}$ ($i,j = 1, \ldots, d_{r}$) of the intertwiner $\mathcal{I}^{r}$, where the bulk line ends on the symmetry (resp. physical) boundary in the $i$-th (resp. $j$-th) way. Using the orthonormality relations in representation theory, it can be proved that the following expression for the vacua fulfils idempotency:
\begin{equation}
\label{eq:vacua-partial-SSB}
    v_{k} = \frac{|H|}{|G|} \sum_{r} d_{r} \sum_{i,j} D_{r}^{ij}(k) \mathcal{I}^{r}_{ji},
\end{equation}
where $k = 1, 2, \ldots, N = \sum_{r}d_{r}^{2} = |G|/|H|$, and $D_{r}^{ij}(k)$ is the representation matrix in $r$. Comparing with the general expression~\eqref{eq:vacua-expression-schematic}, one finds $C_{k,0} = |H|/|G|$, $\forall\, k$. Evidently, the conditional expectation $E_{\mathcal{I}}: v_{k} \mapsto C_{k,0} \mathcal{I}^{(0)}$ is trace-preserving, and the relative entropy 
\begin{equation}
\label{eq:relative-entropy-group}
    S_{\mathcal{I}}(v_{k}|v_{k} \circ E_{\mathcal{I}}) = \log{\left( |G|/|H| \right)}
\end{equation}
is the same for all the symmetry-breaking vacua. Regarding relative Euler terms, the result above establishes that 
\begin{equation}
\label{eq:relative-euler-group}
    \lambda_{k^{\prime},k} = \frac{1}{2} \log\, \left(\frac{C_{k,0}}{C_{k^{\prime},0}}\right)=0\, \quad \forall \, k, k^{\prime}, 
\end{equation}
from which one derives the indistinguishability of the vacua due to the vanishing of all relative Euler terms.

Before proceeding to concrete examples, let us also note that the~\emph{twist operators} $\tau_{[g]}$ are those corresponding to the topological lines $([g],\boldsymbol{1}_{+})$, where $\boldsymbol{1}_{+}$ denotes the trivial representation of $\mathsf{C}_{G}(g)$. The twist operators are dual to the intertwiners in the sense that these topological lines can end on the Neumann boundary with Lagrangian algebra
\begin{equation}
    \mathcal{L}_{\mathrm{Neu}} = \bigoplus_{[g]} ([g],\boldsymbol{1}_{+}).
\end{equation}
The twist operator $\tau_{[g]}$ is invariant under the global $G$ symmetry and lives in the sector twisted by the defect line $[g]$. The Dirichlet boundary condition is converted to Neumann by gauging the full $G$ symmetry. As we shall see in section~\ref{sec:noninvertible}, this further indicates that these twist operators play the role of the intertwiners in the case of a $\mathrm{Rep}(G)$ non-invertible symmetry.

\subsection{Non-anomalous~\texorpdfstring{$\mathbb{Z}_{2}$}{TEXT} symmetry}

Let us begin with the simplest example, where $G = \mathbb{Z}_{2} = \{ \boldsymbol{1}, \eta \}$\footnote{To parallel the notation for the $\mathbb{Z}_{2}$ Tambara-Yamagami fusion category below, here we use $\boldsymbol{1}$ instead of $e$ to denote the unit element.}. The 't Hooft anomaly is specified by an element in $\mathrm{H}^{3}(\mathbb{Z}_{2}, U(1)) = \mathbb{Z}_{2}$. In this subsection, we choose it to be the trivial element. As $\mathbb{Z}_{2}$ is an Abelian group, its (one-dimensional) irreps $\{ +, - \}$ generate the Pontryagin dual that is isomorphic to $\mathbb{Z}_{2}$ itself. Thus, there are four simple topological lines in the DW theory, i.e. the quantum double of $\mathrm{Vec}_{\mathbb{Z}_{2}}$:\footnote{The `electric' line $\mathbf{e}$ is not to be confused with the unit element $e$ of a group. Note also that as $\mathbb{Z}_{2}$ is Abelian, each of its conjugacy classes consists of a single element.}
\begin{equation}
    \mathbf{1} = ([\boldsymbol{1}], +), \quad \mathbf{e} = ([\boldsymbol{1}], -), \quad \mathbf{m} = ([\eta], +), \quad \mathbf{e} \otimes \mathbf{m} = ([\eta], -).
\end{equation}
This SymTFT is nothing but the toric code topological order, where $\mathbf{e}$ and $\mathbf{m}$ are the `electric' and `magnetic' lines, respectively. The symmetry boundary is described by the Lagrangian algebra
\begin{equation}
    \mathcal{L}_{\mathrm{Dir}} = \mathbf{1} \oplus \mathbf{e},
\end{equation}
namely, the electric line can condense on the symmetry boundary; on the other hand, the magnetic line is mapped under the forgetful functor as $\mathbf{m} \mapsto \eta$. In addition to the Dirichlet boundary, the toric code admits another topological boundary condition, namely the Neumann boundary on which the magnetic line condense:
\begin{equation}
    \mathcal{L}_{\mathrm{Neu}} = \mathbf{1} \oplus \mathbf{m}.
\end{equation}
As a result, after compactifying the SymTFT sandwich, one finds two gapped phases for the non-anomalous $\mathbb{Z}_{2}$ symmetry.

Taking the physical boundary condition to be Dirichlet as well yields the $\mathbb{Z}_{2}$ SSB phase; as the bulk line $\mathbf{e}$ can end topologically on both boundaries, one finds a non-trivial intertwiner, which we denote as $\mathcal{I}^{-}$. From~\eqref{eq:vacua-partial-SSB} with $G = \mathbb{Z}_{2}$ and $H = \{ \boldsymbol{1} \}$, the following linear combinations of $\mathcal{I}^{-}$ and the trivial intertwiner $\mathcal{I}^{+}$ (corresponding to the trivial bulk line $\mathbf{1}$) give the vacua satisfying idempotency:
\begin{align}
\label{eq:vacua-Z2-SSB-0}
    v_{\mathbf{1}} &= \frac{1}{2} \left( \mathcal{I}^{+} + \mathcal{I}^{-} \right), \nonumber \\
    v_{\eta} &= \frac{1}{2} \left( \mathcal{I}^{+} - \mathcal{I}^{-} \right).
\end{align}
These vacua form an idempotent basis for the regular representation of $\mathbb{Z}_{2}$. Explicitly, in the basis $\{ v_{\mathbf{1}}, v_{\eta}\}$, the density matrices read
\begin{align}
\rho_{v_{\mathbf{1}}}=\begin{pmatrix}1&0\\0&0\end{pmatrix},\qquad
\rho_{v_\eta}=\begin{pmatrix}0&0\\0&1\end{pmatrix}.
\end{align}
The conditional expectation $E_{\mathcal{I}}$ kills the non-trivial intertwiner and leaves the trivial one invariant. Thus,
\begin{equation}
    v_{\mathbf{1}} \circ E_{\mathcal{I}} = v_{\eta} \circ E_{\mathcal{I}} = \frac{1}{2} \mathcal{I}^{+} = \frac{1}{2} \left( v_{\mathbf{1}} + v_{\eta} \right),
\end{equation}
leading to the symmetrised density matrices
\begin{equation}
    \rho_{v_{\mathbf{1}} \circ E_{\mathcal{I}}} = \rho_{v_{\eta} \circ E_{\mathcal{I}}} = \begin{pmatrix}1/2&0\\0&1/2\end{pmatrix}.
\end{equation}
We see that the vacua become the maximally mixed state, or uniform distribution, under the conditional expectation. Using the definition of the relative entropy~\eqref{eq:relative-entropy-def}, one finds
\begin{equation}
    S_{\mathcal{I}}(v_{\mathbf{1}}|v_{\mathbf{1}} \circ E_{\mathcal{I}}) = S_{\mathcal{I}}(v_{\eta}|v_{\eta} \circ E_{\mathcal{I}}) = \log{2}.
\end{equation}
As expected, the entropic order parameter is identical for both vacua and is equal to $\log{|\mathbb{Z}_{2}|}$.

\begin{figure}[htbp]
\centering
\includegraphics[width=\textwidth]{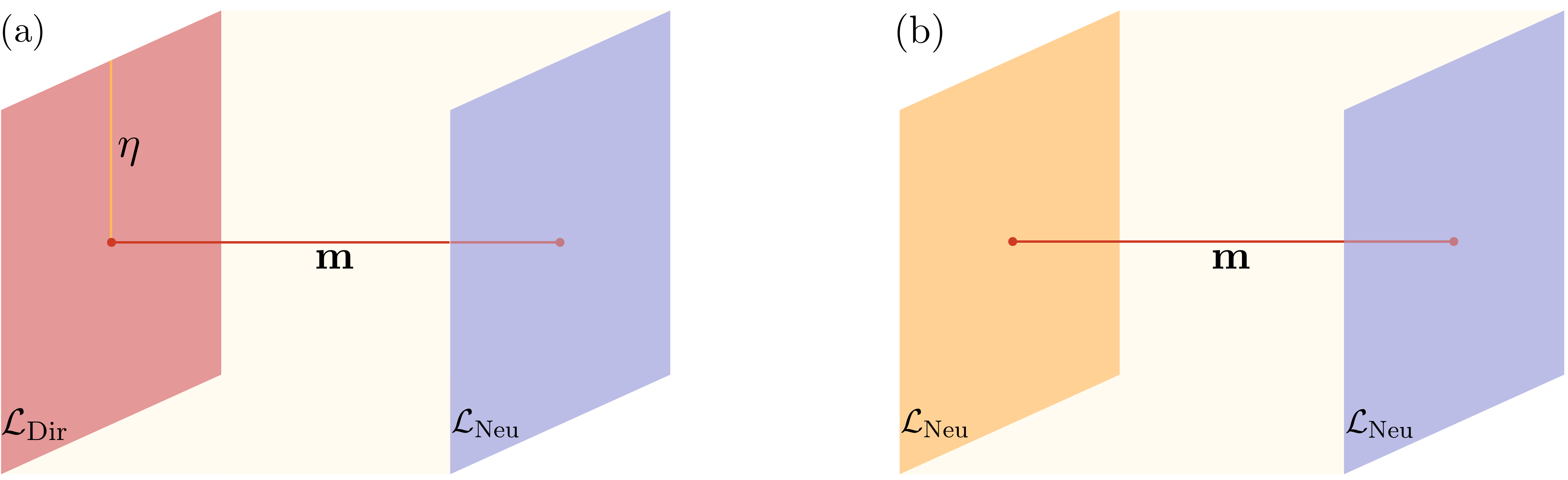}
\caption{(a) The `string order parameter' corresponds to the twisted-sector intertwiner in the $\mathbb{Z}_{2}$ SPT phase. (b) Gauging $\mathbb{Z}_{2}$ on the symmetry boundary converts the Dirichlet boundary condition to Neumann and results in a dual SSB phase, where the string order parameter is mapped to an ordinary one associated with the untwisted intertwiner. \label{fig:5}}
\end{figure}

The other gapped phase for $\mathbb{Z}_{2}$ symmetry is related to the SSB phase via the Kramers-Wannier duality transformation~\cite{kramers1941}, which amounts to switching the physical boundary condition from Dirichlet to Neumann in the SymTFT construction. This produces an SPT phase, as no non-trivial bulk line can end topologically on both boundaries, and there is only one vacuum. In this case, the entropic order parameter defined above vanishes; nevertheless, there exists an intertwiner living in the twisted sector, which corresponds to the bulk topological line $\mathbf{m}$ that condenses on the physical boundary but is mapped to $\eta$ on the symmetry boundary. (It is at the same time a twist operator of the $\mathbb{Z}_{2}$ symmetry.) This is nothing but the `string order parameter' familiar from the SPT literature~\cite{nijs1989}. In fact, it can be mapped to the ordinary (untwisted) order parameter for an SSB phase by gauging the global $\mathbb{Z}_{2}$ symmetry on the symmetry boundary. The situation is easily visualised using the SymTFT graphics, as shown in figure~\ref{fig:5}. The string order parameter is invariant under the action of the $\mathbb{Z}_{2}$ symmetry, indicating that this SPT phase is trivial.

\subsection{Anomalous~\texorpdfstring{$\mathbb{Z}_{2}$}{TEXT} symmetry}

We continue the study of the example and consider $G = \mathbb{Z}_{2}$ with 't Hooft anomaly $\omega \in \mathrm{H}^{3}(\mathbb{Z}_{2}, U(1)) = \mathbb{Z}_{2}$, for which the only non-trivial element reads $\omega(\eta,\eta,\eta) = -1$. The simple topological lines in the DW theory for $\mathrm{Vec}_{\mathbb{Z}_{2}}^{\omega}$ are again labelled by two indices, one of which is the conjugacy class $[g]$ with $g = \mathbf{1}, \eta$. The other index is given by~\emph{twisted} irreps of the centraliser $\mathsf{C}_{G}(g)$, where the twist $\omega_{g} \in \mathrm{H}^{2}(G,U(1))$ is induced by $\omega$ through the slant product~\cite{bhardwaj2025}; in the case of $G = \mathbb{Z}_{2}$, $\mathsf{C}_{G}(g) = G$ and $\omega_{g}$ is trivial. The resulting SymTFT is the double semion model, where the simple topological lines are
\begin{equation}
    \mathbf{1} = ([\boldsymbol{1}], +), \quad \mathbf{s}\bar{\mathbf{s}} = ([\boldsymbol{1}], -), \quad \mathbf{s} = ([\eta], +), \quad \bar{\mathbf{s}} = ([\eta], -),
\end{equation}
where $\mathbf{s}$ is the semion, $\bar{\mathbf{s}}$ is the anti-semion, and their composition $\mathbf{s}\bar{\mathbf{s}}$ is a boson. The cocycle $\omega$ arises in the associating law for these lines. The Dirichlet boundary condition is characterised by the Lagrangian algebra
\begin{equation}
    \mathcal{L}_{\mathrm{Dir}} = \mathbf{1} \oplus \mathbf{s}\bar{\mathbf{s}}.
\end{equation}
Taking both the symmetry boundary and the physical boundary conditions to be Dirichlet, one obtains the $\mathbb{Z}_{2}$ SSB phase. If we denote the intertwiners corresponding to $\mathbf{1}$ and $\mathbf{s}\bar{\mathbf{s}}$ again by 
$\mathcal{I}^{+}$ and $\mathcal{I}^{-}$, respectively, the two vacua are still given by the expression~\eqref{eq:vacua-Z2-SSB-0}, and the relative entropies are the same as those in the non-anomalous case above. However, the~\emph{only} Lagrangian algebra in $\mathcal{Z}(\mathrm{Vec}_{\mathbb{Z}_{2}}^{\omega})$ is the one for the Dirichlet boundary condition; namely, the anomalous $\mathbb{Z}_{2}$ symmetry has to be spontaneously broken, and there is no SPT phase. This reflects the mathematical fact that an anomalous fusion category does not admit a fibre functor.\footnote{In fact, this can be taken as the~\emph{definition} of an anomalous fusion category symmetry. Ref.~\cite{choi2023} studied the relation between this definition and the existence of certain symmetric boundary conditions in (1+1)d QFTs.}

\subsection{\texorpdfstring{$\mathbb{Z}_{2} \times \mathbb{Z}_{2}$}{TEXT} symmetry}

In this subsection, we focus on another example, the (non-anomalous) $\mathbb{Z}_{2} \times \mathbb{Z}_{2}$ symmetry, which is the symmetry pertinent to the celebrated Affleck-Kennedy-Lieb-Tasaki (AKLT) model~\cite{affleck1987}. The latter has been recognised as a prototypical example of non-trivial SPT phases.

We label the non-unit element in the two copies of $\mathbb{Z}_{2}$ by $\eta_{1}$ and $\eta_{2}$, respectively; the Pontryagin dual of $\mathbb{Z}_{2} \times \mathbb{Z}_{2}$ is denoted as $\{ ++,+-,-+,-- \}$ in an obvious manner. The Dirichlet boundary condition is characterised by the Lagrangian algebra
\begin{equation}
    \mathcal{L}_{\mathrm{Dir}} = ([\boldsymbol{1}],++) \oplus ([\boldsymbol{1}],+-) \oplus ([\boldsymbol{1}],-+) \oplus ([\boldsymbol{1}],--).
\end{equation}
Taking both the symmetry and physical boundaries to be Dirichlet produces the gapped phase where the $\mathbb{Z}_{2} \times \mathbb{Z}_{2}$ symmetry is completely broken. After compactifying the SymTFT sandwich, the intertwiners resulted from the above four bulk lines that can end topologically on both boundaries are denoted as $\mathcal{I}^{++}, \mathcal{I}^{+-}, \mathcal{I}^{-+}, \mathcal{I}^{--}$, respectively. Eq.~\eqref{eq:vacua-partial-SSB} then yields the expression for the $\mathbb{Z}_{2} \times \mathbb{Z}_{2}$ SSB vacua that are labelled by the group elements:
\begin{align}
    &v_{\boldsymbol{1}} = \frac{1}{4} \left( \mathcal{I}^{++} + \mathcal{I}^{+-} + \mathcal{I}^{-+} + \mathcal{I}^{--} \right), \nonumber \\
    &v_{\eta_{1}} = \frac{1}{4} \left( \mathcal{I}^{++} + \mathcal{I}^{+-} - \mathcal{I}^{-+} - \mathcal{I}^{--} \right), \nonumber \\
    &v_{\eta_{2}} = \frac{1}{4} \left( \mathcal{I}^{++} - \mathcal{I}^{+-} + \mathcal{I}^{-+} - \mathcal{I}^{--} \right), \nonumber \\
    &v_{\eta_{1}\eta_{2}} = \frac{1}{4} \left( \mathcal{I}^{++} - \mathcal{I}^{+-} - \mathcal{I}^{-+} + \mathcal{I}^{--} \right).
\end{align}
Again, the entropic order parameter
\begin{equation}
    S_{\mathcal{I}}(v_{g}|v_{g} \circ E_{\mathcal{I}}) = \log{|\mathbb{Z}_{2} \times \mathbb{Z}_{2}|} = \log{4}
\end{equation}
is the same for the four vacua.

There are three $\mathbb{Z}_{2}$ subgroups (the two copies of $\mathbb{Z}_{2}$, and the diagonal subgroup) of $\mathbb{Z}_{2} \times \mathbb{Z}_{2}$ that can be gauged on the physical boundary. Correspondingly, there are three Lagrangian algebras
\begin{align}
    &\mathcal{L}_{\mathrm{Neu}(\mathbb{Z}_{2}),1} = ([\boldsymbol{1}],++) \oplus ([\boldsymbol{1}],+-) \oplus ([\eta_{1}],++) \oplus ([\eta_{1}],+-), \nonumber \\
    &\mathcal{L}_{\mathrm{Neu}(\mathbb{Z}_{2}),2} = ([\boldsymbol{1}],++) \oplus ([\eta_{2}],++) \oplus ([\boldsymbol{1}],-+) \oplus ([\eta_{2}],-+), \nonumber \\
    &\mathcal{L}_{\mathrm{Neu}(\mathbb{Z}_{2}),3} = ([\boldsymbol{1}],++) \oplus ([\eta_{1}\eta_{2}],++) \oplus ([\eta_{1}\eta_{2}],--) \oplus ([\boldsymbol{1}],--),
\end{align}
which lead to three gapped $\mathbb{Z}_{2}$ SSB phases. Thanks to the obvious invariance under permutation, these phases share the same properties; without loss of generality, we focus on $\mathcal{L}_{\mathrm{Neu}(\mathbb{Z}_{2}),1}$. The intertwiners produced by bulk topological lines that can end on both boundaries are $\mathcal{I}^{++}$ and $\mathcal{I}^{+-}$, and the resulting vacua read
\begin{align}
    &v_{\boldsymbol{1}} = \frac{1}{2} \left( \mathcal{I}^{++} + \mathcal{I}^{+-} \right), \nonumber \\
    &v_{\eta_{2}} = \frac{1}{2} \left( \mathcal{I}^{++} - \mathcal{I}^{+-} \right).
\end{align}
Note that we use $\{ \boldsymbol{1}, \eta_{2} \}$ to label these vacua to emphasise that the~\emph{broken} symmetry is the second copy of $\mathbb{Z}_{2}$. As expected, the relative entropy of both vacua is $\log{|\mathbb{Z}_{2}|} = \log{2}$.

Gauging the full $\mathbb{Z}_{2} \times \mathbb{Z}_{2}$ symmetry on the physical boundary amounts to choosing the Neumann boundary condition
\begin{equation}
    \mathcal{L}_{\mathrm{Neu}(\mathbb{Z}_{2} \times \mathbb{Z}_{2})} = ([\boldsymbol{1}],++) \oplus ([\eta_{1}],++) \oplus ([\eta_{2}],++) \oplus ([\eta_{1}\eta_{2}],++).
\end{equation}
The resulting phase is the trivial symmetric phase, where the string order parameters are invariant under the $\mathbb{Z}_{2} \times \mathbb{Z}_{2}$ symmetry. A crucial difference compared to the case of $\mathbb{Z}_{2}$ case is that $\mathrm{H}^{2}(\mathbb{Z}_{2} \times \mathbb{Z}_{2}, U(1)) = \mathbb{Z}_{2}$; this means that one can also perform the gauging with a discrete torsion given by the non-identity element in $\mathrm{H}^{2}(\mathbb{Z}_{2} \times \mathbb{Z}_{2}, U(1))$. The corresponding Lagrangian algebra for the physical boundary is
\begin{equation}
    \mathcal{L}_{\mathrm{Neu}(\mathbb{Z}_{2} \times \mathbb{Z}_{2})}^{\prime} = ([\boldsymbol{1}],++) \oplus ([\eta_{1}],+-) \oplus ([\eta_{2}],-+) \oplus ([\eta_{1}\eta_{2}],--).
\end{equation}
Again, the resulting phase has a single vacuum. However, this is a~\emph{non-trivial} SPT phase because of the non-trivial commutation relations between the intertwiners associated with the bulk topological lines $([\eta_{1}],+-), ([\eta_{2}],-+), ([\eta_{1}\eta_{2}],--)$ and the twist operators for the $\mathbb{Z}_{2} \times \mathbb{Z}_{2}$ symmetry. In fact, it is nothing but the SPT phase represented by the AKLT state. The analogue of the duality transformation shown in figure~\ref{fig:5} that relates the AKLT phase to an SSB phase is known as the Kennedy-Tasaki transformation~\cite{kennedy1992a,oshikawa1992,kennedy1992b}, which amounts to `twisted gauging' $\mathbb{Z}_{2} \times \mathbb{Z}_{2}$~\cite{li2023} on the symmetry boundary of the SymTFT.

\subsection{\texorpdfstring{$S_{3}$}{TEXT} symmetry}

The simplest non-Abelian group is $S_{3}$, the permutation group of three objects. Let us focus on the example $G = S_{3}$ in this subsection.

The $6$ elements of $S_{3}$ are denoted as
\begin{equation}
    S_{3} = \{ e, a, a^{2}, b, ab, a^{2}b \},
\end{equation}
where $a$ generates a cyclic permutation, satisfying
\begin{equation}
    a^{3} = e,
\end{equation}
and $b$ generates a swapping, satisfying
\begin{equation}
    b^{2} = e.
\end{equation}
$a$ and $b$ have non-trivial commutation relation,
\begin{equation}
    ba = a^{2}b \neq ba^{2} = ab.
\end{equation}
There are $3$ conjugacy classes:
\begin{equation}
    [e] = \{ e \}, \quad [a] = \{ a, a^{2} \}, \quad [b] = \{ b, ab, a^{2}b \};
\end{equation}
the corresponding centralisers are
\begin{equation}
    \mathsf{C}_{S_{3}}(e) = S_{3}, \quad \mathsf{C}_{S_{3}}(a) = \{ e, a, a^{2}\} = \mathbb{Z}_{3}, \quad \mathsf{C}_{S_{3}}(b) = \{ e, b \} = \mathbb{Z}_{2}.
\end{equation}
$S_{3}$ has three irreps: the trivial representation $\boldsymbol{1}_{+}$, the sign representation $\boldsymbol{1}_{-}$, and the two-dimensional representation $\boldsymbol{2}$. The decomposition rules of the tensor products among these irreps read
\begin{equation}
    \boldsymbol{1}_{-} \otimes \boldsymbol{1}_{-} = \boldsymbol{1}_{+}, \quad \boldsymbol{1}_{-} \otimes \boldsymbol{2} = \boldsymbol{2}, \quad \boldsymbol{2} \otimes \boldsymbol{2} = \boldsymbol{1}_{+} \oplus \boldsymbol{1}_{-} \oplus \boldsymbol{2}.
\end{equation}

In the case where the $S_{3}$ symmetry is completely broken, both the symmetry and the physical boundary conditions of the SymTFT are taken to be Dirichlet. The associated Lagrangian algebra reads
\begin{equation}
\label{eq:Dir-S3}
    \mathcal{L}_{\mathrm{Dir}} = ([e],\boldsymbol{1}_{+}) \oplus ([e],\boldsymbol{1}_{-}) \oplus 2([e],\boldsymbol{2}) = \bigoplus_{r} d_{r} ([e],r).
\end{equation}
This is nothing but the group algebra of $G = S_{3}$ that serves as the representation space of the regular representation, where the number of times an irrep appears is equal to its dimension. From~\eqref{eq:Dir-S3}, the bulk topological line of the form $([e],r)$ can end topologically on both symmetry and physical boundaries in $d_{r}$ different ways; these correspond precisely to the components $\mathcal{I}^{r}_{ij}$ ($i,j = 1, \ldots, d_{r}$) of the intertwiner $\mathcal{I}^{r}$, where the bulk line ends on the symmetry (resp. physical) boundary in the $i$-th (resp. $j$-th) way. For the complete breaking of a finite symmetry group $G$, the expression~\eqref{eq:vacua-partial-SSB} of the $N = \sum_{r}d_{r}^{2} = |G|$ symmetry-broken vacua reduces to
\begin{equation}
\label{eq:vacua-S3-SSB}
    v_{g} = \sum_{r} \frac{d_{r}}{|G|} \sum_{i,j} D_{r}^{ij}(g) \mathcal{I}^{r}_{ji},
\end{equation}
where $g$ labels the elements of $G$. One has $C_{g,0} = |G|^{-1}$, and the relative entropy $S_{\mathcal{I}}(v_{g}|v_{g} \circ E_{\mathcal{I}}) = \log{|G|}$ is the same for any $v_{g}$. For $G = S_{3}$, the six vacua explicitly read
\begin{align}
    &v_{e} = \frac{1}{6} \left( \mathcal{I}^{\boldsymbol{1}_{+}} + \mathcal{I}^{\boldsymbol{1}_{-}} + 2\mathcal{I}^{\boldsymbol{2}}_{11} + 2\mathcal{I}^{\boldsymbol{2}}_{22} \right), \nonumber \\
    &v_{a} = \frac{1}{6} \left( \mathcal{I}^{\boldsymbol{1}_{+}} + \mathcal{I}^{\boldsymbol{1}_{-}} + 2\omega\mathcal{I}^{\boldsymbol{2}}_{11} + 2\omega^{2}\mathcal{I}^{\boldsymbol{2}}_{22} \right), \nonumber \\
    &v_{a^{2}} = \frac{1}{6} \left( \mathcal{I}^{\boldsymbol{1}_{+}} + \mathcal{I}^{\boldsymbol{1}_{-}} + 2\omega^{2}\mathcal{I}^{\boldsymbol{2}}_{11} + 2\omega\mathcal{I}^{\boldsymbol{2}}_{22} \right), \nonumber \\
    &v_{b} = \frac{1}{6} \left( \mathcal{I}^{\boldsymbol{1}_{+}} - \mathcal{I}^{\boldsymbol{1}_{-}} + 2\mathcal{I}^{\boldsymbol{2}}_{12} + 2\mathcal{I}^{\boldsymbol{2}}_{21} \right), \nonumber \\
    &v_{ba} = \frac{1}{6} \left( \mathcal{I}^{\boldsymbol{1}_{+}} - \mathcal{I}^{\boldsymbol{1}_{-}} + 2\omega\mathcal{I}^{\boldsymbol{2}}_{12} + 2\omega^{2}\mathcal{I}^{\boldsymbol{2}}_{21} \right), \nonumber \\
    &v_{ab} = \frac{1}{6} \left( \mathcal{I}^{\boldsymbol{1}_{+}} - \mathcal{I}^{\boldsymbol{1}_{-}} + 2\omega^{2}\mathcal{I}^{\boldsymbol{2}}_{12} + 2\omega\mathcal{I}^{\boldsymbol{2}}_{21} \right),
\end{align}
where $\omega = \mathrm{e}^{2\pi\mathrm{i}/3}$.

The case where a (non-anomalous) subgroup $H$ of $G$ remains unbroken can be studied analogously. For $G = S_{3}$, we can choose Neumann boundary condition for $H = \mathsf{C}_{S_{3}}(b) = \{ e, b \} = \mathbb{Z}_{2}$, for which the Lagrangian algebra characterising the physical boundary is
\begin{equation}
    \mathcal{L}_{\text{Neu}(\mathbb{Z}_{2})} = ([e],\boldsymbol{1}_{+}) \oplus ([e],\boldsymbol{2}) \oplus ([b],\boldsymbol{1}_{+}),
\end{equation}
where the $\boldsymbol{1}_{+}$ in $([b],\boldsymbol{1}_{+})$ denotes the trivial representation of $\mathsf{C}_{S_{3}}(b) = \mathbb{Z}_{2}$. In the same vein as~\eqref{eq:vacua-S3-SSB}, the vacua are obtained by linear combinations of the intertwiners $\mathcal{I}^{\boldsymbol{1}_{+}}$ and $\mathcal{I}^{\boldsymbol{2}}_{i}$ with $i = 1, 2$ [as $([e],\boldsymbol{2})$ can end in two different ways on the symmetry boundary but only in one way on the physical boundary]. The result reads
\begin{align}
\label{eq:vacua-Z3-SSB}
    &v_{e} = \frac{1}{3} \left( \mathcal{I}^{\boldsymbol{1}_{+}} + \mathcal{I}^{\boldsymbol{2}}_{1} + \mathcal{I}^{\boldsymbol{2}}_{2}\right), \nonumber \\
    &v_{a} = \frac{1}{3} \left( \mathcal{I}^{\boldsymbol{1}_{+}} + \omega \mathcal{I}^{\boldsymbol{2}}_{1} + \omega^{2} \mathcal{I}^{\boldsymbol{2}}_{2}\right), \nonumber \\
    &v_{a^{2}} = \frac{1}{3} \left( \mathcal{I}^{\boldsymbol{1}_{+}} + \omega^{2} \mathcal{I}^{\boldsymbol{2}}_{1} + \omega \mathcal{I}^{\boldsymbol{2}}_{2}\right).
\end{align}
Here, $\{ e, a, a^{2} \}$ form the group $\mathbb{Z}_{3}$ representing the symmetry that is~\emph{broken}, and the two-dimensional irrep $\boldsymbol{2}$ for $S_{3}$ decomposes as the direct sum of one-dimensional irreps $\boldsymbol{1}_{\omega}$ and $\boldsymbol{1}_{\omega^{2}}$ for $\mathbb{Z}_{3}$. We can also consider the case of $H = \mathsf{C}_{S_{3}}(a) = \{ e, a, a^{2}\} = \mathbb{Z}_{3}$ as the unbroken symmetry. The Lagrangian algebra for the physical boundary reads
\begin{equation}
    \mathcal{L}_{\text{Neu}(\mathbb{Z}_{3})} = ([e],\boldsymbol{1}_{+}) \oplus ([e],\boldsymbol{1}_{-}) \oplus 2([a],\boldsymbol{1}_{1}),
\end{equation}
where $\boldsymbol{1}_{1}$ stands for the trivial representation of $\mathbb{Z}_{3}$. Clearly, there are only two independent single-component intertwiners $\mathcal{I}^{\boldsymbol{1}_{+}}$ and $\mathcal{I}^{\boldsymbol{1}_{-}}$ that can end on both topological boundaries, and the vacua are simply given by
\begin{align}
\label{eq:vacua-Z2-SSB}
    &v_{e} = \frac{1}{2} \left( \mathcal{I}^{\boldsymbol{1}_{+}} + \mathcal{I}^{\boldsymbol{1}_{-}} \right), \nonumber \\
    &v_{b} = \frac{1}{2} \left( \mathcal{I}^{\boldsymbol{1}_{+}} - \mathcal{I}^{\boldsymbol{1}_{-}} \right),
\end{align}
where $\{ e, b \}$ form the broken $\mathbb{Z}_{2}$ symmetry. For all the above cases, the relative entropy is the same for each symmetry-broken vacuum and agrees with the general result $S_{\mathcal{I}}(v_{k}|v_{k} \circ E_{\mathcal{I}}) = \log{\left( |G|/|H| \right)}$.

Finally, one can take the Lagrangian algebra corresponding to the physical boundary to be
\begin{equation}
\label{eq:Neu-S3}
    \mathcal{L}_{\mathrm{Neu}(S_{3})} = ([e],\boldsymbol{1}_{+}) \oplus ([a],\boldsymbol{1}_{1}) \oplus ([b],\boldsymbol{1}_{+}),
\end{equation}
where the $\boldsymbol{1}_{1}$ in $([a],\boldsymbol{1}_{1})$ denotes the trivial representation of $\mathsf{C}_{S_{3}}(a) = \mathbb{Z}_{3}$, and the $\boldsymbol{1}_{+}$ in $([b],\boldsymbol{1}_{+})$ the trivial representation of $\mathsf{C}_{S_{3}}(b) = \mathbb{Z}_{2}$. The only bulk line that can end topologically on both boundaries is $([e],\boldsymbol{1}_{+})$; this is the trivial (SPT) phase for the $S_3$ symmetry.

\section{Non-invertible symmetries}
\label{sec:noninvertible}

Let us turn to the more interesting case of non-invertible symmetries. As discussed above, the components of the (untwisted) intertwiners are in correspondence with the simple objects of a module category over the symmetry category. For simplicity, we will consider the situations where these intertwiners themselves form a modular category $\mathcal{M}$. In these situations, one can conveniently use the Verlinde formula~\cite{verlinde1988} to show that the expression for the vacua,
\begin{equation}
\label{eq:vacua-expression-non-invertible}
    v_{x} = \mathbb{S}_{\boldsymbol{1}x}^{*} \sum_{y} \mathbb{S}_{yx} \mathcal{I}^{y},
\end{equation}
fulfils idempotency, where $x, y, \ldots$ label the simple objects, $\boldsymbol{1}$ is the identity object, and $\mathbb{S}$ is the modular S matrix. The conditional expectation acts as 
\begin{equation}
    E: v_{x} \mapsto |\mathbb{S}_{\boldsymbol{1}x}|^{2}\cdot\mathcal{I}^{\boldsymbol{1}} = \frac{d_{x}^{2}}{\mathrm{dim}(\mathcal{M})} \cdot\mathcal{I}^{\boldsymbol{1}},
\end{equation}
where $d_{x}$ is the quantum dimension of the simple object $x$, and $\mathrm{dim}(\mathcal{M}) = \sum_{x} d_{x}^{2}$ is the total quantum dimension of $\mathcal{M}$. The density matrix of the symmetrised state is 
\begin{equation}
    \rho_{v_{x} \circ E} = \frac{d_{x}^{2}}{\mathrm{dim}(\mathcal{M})}\cdot\mathbb{I}.
\end{equation}
 One observes that $E$ is not necessarily trace-preserving in this situation. To compute the entropic order parameter, one has to go back to the definition~\eqref{eq:relative-entropy-def} of a relative entropy; the result is
\begin{equation}
\label{eq:relative-entropy-general}
    S(v_{x} | v_{x} \circ E) = \log{\frac{\mathrm{dim}(\mathcal{M})}{d_{x}^{2}}},
\end{equation}
which in general depends on the vacuum $v_{x}$, since the quantum dimensions associated with different vacua can differ. This implies that the vacua can be~\emph{distinguishable}. For a group symmetry, on the other hand, all quantum dimensions are equal to $1$, leading to the same relative entropy~\eqref{eq:relative-entropy-group}. With regard to relative Euler terms, the result above implies that 
\begin{equation}
\label{eq:relative-euler-cat}
   \lambda_{x,y} = \frac{1}{2} \log\, \left(\frac{C_{y,0}}{C_{x,0}}\right)=\log\, \frac{d_y}{d_x}\,,
\end{equation}
which are not all necessarily vanishing as in the case of group symmetries \eqref{eq:relative-euler-group}.

The representation category $\mathrm{Rep}(G)$ of a finite group $G$ has the same Drinfeld centre as $\mathrm{Vec}_{G}$; for both global symmetries $G$ and $\mathrm{Rep}(G)$, the SymTFT is the $G$-DW gauge theory. In fact, these two symmetries are related to each other by (re-)gauging and are mutually dual in the `electric-magnetic' sense, and the Dirichlet boundary condition for $\mathrm{Rep}(G)$ is equivalent to the Neumann boundary condition for $G$; namely, the symmetry boundary of the SymTFT for $\mathrm{Rep}(G)$ corresponds to the Lagrangian algebra $\mathcal{L}_{\mathrm{Neu}(G)}$. This duality is an example of more general~\emph{Morita equivalent} fusion categories, which share the same Drinfeld centre. Therefore, the topological holography approach is particularly helpful in studying Morita equivalent symmetry categories, which, by definition, share a common SymTFT that encodes the duality data relating them. As we mentioned in section~\ref{sec:invertible}, for the symmetry $\mathrm{Rep}(G)$, the role of intertwiners is played by the twist operators for a subgroup $H$ of $G$, when the physical boundary is taken as the Neumann boundary condition for $H$. These twist operators $\tau_{[h]}$ are labelled by the conjugacy classes in $H$. Note that for a finite group, the number of conjugacy classes is equal to that of irreps; in~\eqref{eq:vacua-expression-non-invertible}, the S matrix elements are given in terms of the characters of the irreps evaluated on the conjugacy classes. The vacua are labelled by the irreps of $H$:
\begin{equation}
\label{eq:vacua-twists}
    v_{r} = \frac{d_{r}}{|H|} \sum_{[h]} \chi_{r}([h]) \tau_{[h]}.
\end{equation}
The result for the relative entropies,
\begin{equation}
\label{eq:relative-entropy-Rep}
    S_{\tau}(v_{r} | v_{r} \circ E_{\tau}) = \log{\frac{|H|}{d_{r}^{2}}},
\end{equation}
where we have used the fact that $\chi_{r}([e]) = d_{r}$ is the dimension of the irrep $r$, is simply a special case of the general expression~\eqref{eq:relative-entropy-general}. For a non-Abelian $H$, the vacua acquire different relative entropies because the irreps of a non-Abelian group have different dimensions. In the following, we consider the group-theoretical non-invertible symmetry $\mathrm{Rep}(S_{3})$ and the Ising fusion category symmetry that is beyond group-theoretical.

\subsection{\texorpdfstring{$\mathrm{Rep}(S_{3})$}{TEXT} symmetry}
First, we focus on $\mathrm{Rep}(S_{3})$, the fusion category formed by representations of the group $S_{3}$, as a global symmetry. 

For the gapped phase with $\mathrm{Rep}(S_{3})$ completely broken, the physical boundary is chosen to be the same as the symmetry boundary, i.e., the topological boundary condition specified by $\mathcal{L}_{\mathrm{Neu}(S_{3})}$ in~\eqref{eq:Neu-S3}. The bulk lines that can end topologically on both boundaries are then identified with the twist operators $\tau_{[g]}$ for $g = e, a, b$. The vacua are given by~\eqref{eq:vacua-twists} with $H = G = S_{3}$:
\begin{align}
    v_{\boldsymbol{1}_{+}} &= \frac{1}{6} \left( \tau_{[e]} + \tau_{[a]} + \tau_{[b]} \right), \nonumber \\
    v_{\boldsymbol{1}_{-}} &= \frac{1}{6} \left( \tau_{[e]} + \tau_{[a]} - \tau_{[b]} \right), \nonumber \\
    v_{\boldsymbol{2}} &= \frac{1}{3} \left( 2\tau_{[e]} - \tau_{[a]} \right).
\end{align}
The conditional expectation $E_{\tau}$ kills all the non-trivial twist operators, and the resulting relative entropies are
\begin{equation}
    S_{\mathcal{\tau}}(v_{\boldsymbol{1}_{+}}|v_{\boldsymbol{1}_{+}} \circ E_{\tau}) = S_{\mathcal{\tau}}(v_{\boldsymbol{1}_{-}}|v_{\boldsymbol{1}_{-}} \circ E_{\tau}) = \log{6}, \quad S_{\mathcal{\tau}}(v_{\boldsymbol{2}}|v_{\boldsymbol{2}} \circ E_{\tau}) = \log{\frac{3}{2}}.
\end{equation}
We find the relative entropy for $v_{\boldsymbol{2}}$ different from that for $v_{\boldsymbol{1}_{+}}$ and $v_{\boldsymbol{1}_{-}}$, whilst the latter two have the same entropic order parameter. Physically, this indicates that $v_{\boldsymbol{1}_{+}}$ and $v_{\boldsymbol{1}_{-}}$ are not distinguishable from each other, but they are distinguishable from $v_{\boldsymbol{2}}$. Equivalently, the relative Euler terms allow to distinguish the vacua as 
\begin{equation}
    \lambda_{\boldsymbol{1}_{+},\boldsymbol{1}_{-}}=0\,, \quad \lambda_{\boldsymbol{1}_{\pm}, \boldsymbol{2}}=\log 2\, .
\end{equation}

The physical boundary can also be chosen to be $\mathcal{L}_{\mathrm{Neu}(\mathbb{Z}_{2})}$. The bulk lines that can end topologically on both boundaries are now $([e],\boldsymbol{1}_{+})$ and $([b],\boldsymbol{1}_{+})$. Here, $[b]$ can be viewed as the (only non-trivial) irrep of $\mathrm{Rep}(\mathbb{Z}_{2}) \simeq \mathbb{Z}_{2}$ (in the dual sense). Replacing $H$ by $\mathbb{Z}_{2}$ in~\eqref{eq:vacua-twists}, one immediately gets
\begin{align}
\label{eq:vacua-RepZ2-SSB}
    &v_{\boldsymbol{1}_{+}} = \frac{1}{2} \left( \tau_{[e]} + \tau_{[b]} \right), \nonumber \\
    &v_{\boldsymbol{1}_{-}} = \frac{1}{2} \left( \tau_{[e]} - \tau_{[b]} \right).
\end{align}
This is the phase where the sub-symmetry $\mathbb{Z}_{2}$ is broken. Analogously, choosing the physical boundary to be $\mathcal{L}_{\text{Neu}(\mathbb{Z}_{3})}$, the bulk lines that can end topologically on both boundaries are $([e],\boldsymbol{1}_{+})$ and $([a],\boldsymbol{1}_{1})$. A complication for this case is that the line $([a],\boldsymbol{1}_{1})$ can end in two different ways on the physical boundary. This is consistent with the fact that $[a]$, as a two-dimensional representation of $\mathrm{Rep}(\mathbb{Z}_{3}) \simeq \mathbb{Z}_{3}$, is reducible and decomposes into two irreducible ones; the resulting components of the twist operator are denoted as $\tau_{[a],1}$ and $\tau_{[a],2}$. The vacua follow from~\eqref{eq:vacua-twists} as
\begin{align}
\label{eq:vacua-RepZ3-SSB}
    &v_{\boldsymbol{1}_{1}} = \frac{1}{3} \left( \tau_{[e]} + \tau_{[a],1} + \tau_{[a],2} \right), \nonumber \\
    &v_{\boldsymbol{1}_{\omega}} = \frac{1}{3} \left( \tau_{[e]} + \omega \tau_{[a],1} + \omega^{2} \tau_{[a],2} \right), \nonumber \\
    &v_{\boldsymbol{1}_{\omega^{2}}} = \frac{1}{3} \left( \tau_{[e]} + \omega^{2} \tau_{[a],1} + \omega \tau_{[a],2} \right).
\end{align}
This phase is termed the $\mathrm{Rep}(S_{3})/\mathbb{Z}_{2}$ breaking phase in refs.~\cite{bhardwaj2025a,bhardwaj2024}, as the sub-symmetry $\mathbb{Z}_{2}$ is~\emph{unbroken}. It is clear that the $\mathrm{Rep}(S_{3})$ phases~\eqref{eq:vacua-RepZ2-SSB} and~\eqref{eq:vacua-RepZ3-SSB} are dual to the $S_3$ phases~\eqref{eq:vacua-Z2-SSB} and~\eqref{eq:vacua-Z3-SSB}, respectively. For these phases, the entropic order parameters of the symmetry-broken vacua are the same, indicating that the latter are indistinguishable. Finally, taking the physical boundary to be $\mathcal{L}_{\text{Dir}(S_{3})}$ yields the trivial (SPT) phase for the $\mathrm{Rep}(S_{3})$ symmetry.

\subsection{\texorpdfstring{Ising}{TEXT} symmetry}

The case of $\mathrm{Rep}(G)$ considered above belongs to the so-called~\emph{group-theoretical} symmetries. We now go a step further and consider non-group-theoretical (or `intrinsic') non-invertible symmetries, for which the simplest example is given by the $\mathbb{Z}_{2}$ Tambara-Yamagami fusion category $\mathrm{TY}(\mathbb{Z}_{2})$ (with trivial bicharacter and Frobenius-Schur indicator)~\cite{tambara1998}. The simple objects in this fusion category famously describe the topological lines in the Ising CFT~\cite{chang2019}: the trivial line $\boldsymbol{1}$, the line of spin-flip $\eta$, and that of the Kramers-Wannier self-duality $\mathcal{N}$, which satisfy the non-trivial fusion rules
\begin{equation}
    \eta \otimes \eta = \boldsymbol{1}, \quad \eta \otimes \mathcal{N} = \mathcal{N}, \quad \mathcal{N} \otimes \mathcal{N} = \boldsymbol{1} \oplus \eta.
\end{equation}
For this reason, this symmetry category is also denoted simply as $\mathrm{Ising}$. As the latter is a modular category, its Drinfeld centre is given by $\mathrm{Ising} \boxtimes \overline{\mathrm{Ising}}$ describing the `double Ising' topological quantum field theory, where $\overline{\mathrm{Ising}}$ is the orientation reversal\footnote{The twist factors in $\overline{\mathrm{Ising}}$ are the complex conjugate of those in $\mathrm{Ising}$, whereas the remaining data are the same.} of $\mathrm{Ising}$. There are nine simple topological lines labelling by pairs $(x,y)$ in the SymTFT with $x$ (resp. $y$) a simple object of $\mathrm{Ising}$ (resp. $\overline{\mathrm{Ising}}$), and the bulk-to-boundary map is simply the fusion product: $(x,y) \mapsto x \otimes y$. It turns out that there is only one Lagrangian algebra in the Drinfeld centre, which corresponds to the symmetry boundary
\begin{equation}
    \mathcal{L}_{\mathrm{Dir}} = (\boldsymbol{1},\boldsymbol{1}) \oplus (\eta,\eta) \oplus (\mathcal{N},\mathcal{N}).
\end{equation}
Therefore, there is only one gapped phase for the Ising symmetry, which is obtained by choosing both the symmetry and physical boundaries to be $\mathcal{L}_{\mathrm{Dir}}$. This is the phase where the Ising symmetry is completely broken. Corresponding to the bulk lines that can end topologically on both boundaries, there are intertwiners $\mathcal{I}^{(\boldsymbol{1},\boldsymbol{1})}, \mathcal{I}^{(\eta,\eta)}, \mathcal{I}^{(\mathcal{N},\mathcal{N})}$; the latter generate the `diagonal' part of the fusion algebra of the Drinfeld centre, which coincides with that of $\mathrm{Ising}$. In other words, here $\mathcal{M} = \mathrm{Ising}$ is the regular module category over $\mathrm{Ising}$ itself. Therefore, the general expression~\eqref{eq:vacua-expression-non-invertible} for the vacua becomes\footnote{Notice its close formal similarity with the construction of Cardy states in terms of Ishibashi states in rational CFTs~\cite{kishimoto2004}.}
\begin{equation}
    v_{x} = \mathbb{S}_{\boldsymbol{1}x}^{*} \sum_{y} \mathbb{S}_{yx} \mathcal{I}^{(y,y)},
\end{equation}
where $x, y = \boldsymbol{1}, \eta, \mathcal{N}$, and the $\mathrm{Ising}$ modular S matrix reads (the rows and columns are arranged in the order $\boldsymbol{1}, \eta, \mathcal{N}$)
\begin{equation}
    \mathbb{S} = \frac{1}{2} \left( \begin{array}{ccc}
        1  &~ 1 &~ \sqrt{2}\\
        1 &~ 1 &~ -\sqrt{2}\\
        \sqrt{2} &~ -\sqrt{2} &~ 0
    \end{array} \right).
\end{equation}
More explicitly, one has
\begin{align}
\label{eq:vacua-Ising-SSB}
    &v_{\boldsymbol{1}} = \frac{1}{4} \left( \mathcal{I}^{(\boldsymbol{1},\boldsymbol{1})} + \mathcal{I}^{(\eta,\eta)} + \sqrt{2}\mathcal{I}^{(\mathcal{N},\mathcal{N})} \right), \nonumber \\
    &v_{\eta} = \frac{1}{4} \left( \mathcal{I}^{(\boldsymbol{1},\boldsymbol{1})} + \mathcal{I}^{(\eta,\eta)} - \sqrt{2}\mathcal{I}^{(\mathcal{N},\mathcal{N})} \right), \nonumber \\
    &v_{\mathcal{N}} = \frac{1}{2} \left( \mathcal{I}^{(\boldsymbol{1},\boldsymbol{1})} - \mathcal{I}^{(\eta,\eta)} \right).
\end{align}
The conditional expectation $E_{\mathcal{I}}$ kills the non-trivial intertwiners $\mathcal{I}^{(\eta,\eta)}$ and $\mathcal{I}^{(\mathcal{N},\mathcal{N})}$, yielding the relative entropies
\begin{equation}
    S_{\mathcal{I}}(v_{\boldsymbol{1}}|v_{\boldsymbol{1}} \circ E_{\mathcal{I}}) = S_{\mathcal{I}}(v_{\eta}|v_{\eta} \circ E_{\mathcal{I}}) = \log{4}, \quad S_{\mathcal{I}}(v_{\mathcal{N}}|v_{\mathcal{N}} \circ E_{\mathcal{I}}) = \log{2}.
\end{equation}
With $d_{\boldsymbol{1}} = d_{\eta} = 1, d_{\mathcal{N}} = \sqrt{2}$ and $\mathrm{dim}(\mathrm{Ising}) = 4$, it is evident that this result agrees with the general formula~\eqref{eq:relative-entropy-general}. Therefore, the relative Euler terms allow to distinguish the vacua as 
\begin{equation}
    \lambda_{\boldsymbol{1},\eta}=0\,, \quad \lambda_{\boldsymbol{1},\, \mathcal{N}} =\lambda_{\eta,\, \mathcal{N}}=\frac{1}{2}\log 2\, .
\end{equation}
Again, the distinguishability between $v_{\boldsymbol{1}}$ or $v_{\eta}$ and $v_{\mathcal{N}}$ is attributed to the difference in quantum dimensions of the corresponding simple objects.

\section{Conclusions}

In this work, we have shown that topological holography (or SymTFT) and entropic order parameters provide a natural and unified framework for characterizing spontaneous symmetry breaking of generalized symmetries in $(1+1)$ dimensions. In SymTFT, symmetry-broken vacua arise as orthogonal idempotents built from linear combinations of intertwiners associated with bulk topological lines, whilst relative entropy quantifies the loss of information induced by restricting to the symmetry-invariant operator algebra. This makes the origin of the entropic order parameter particularly transparent, and places the spontaneous breaking of invertible and non-invertible symmetries on equal footing. The computation of entropic order parameters is illustrated explicitly for several examples: $\mathbb{Z}_{2}$ (possibly anomalous), $\mathbb{Z}_{2} \times \mathbb{Z}_{2}$, $S_{3}$, $\mathrm{Rep}(S_{3})$, and $\mathrm{Ising}$.

For non-invertible symmetries, this approach also yields a clear information-theoretic understanding of the distinguishability of symmetry-broken vacua. In the examples based on \(\mathrm{Rep}(S_3)\) and the Ising fusion category, the entropic order parameters are controlled by the quantum dimensions of the simple objects labelling the vacua, leading in general to non-vanishing relative Euler terms, which are related to ratios of the quantum dimensions. Thus, unlike the case of ordinary group symmetries, spontaneously broken non-invertible symmetries can produce vacua that are intrinsically distinguishable already at the level of the observable algebra. These results underscore the utility of entropic order parameters, constructed from information-theoretic quantities such as relative entropies, as diagnostics of generalized symmetry breaking, with SymTFT furnishing an efficient framework for organizing the kinematics of gapped phases in $(1+1)$ dimensions.

\acknowledgments

H.-C.Z. acknowledges support by an appointment to the Young Scientist Training (YST) Program at Asia Pacific Center for Theoretical Physics (APCTP) through the Science and Technology Promotion Fund and Lottery Fund of the Korean Government, and support by the Korean Local Governments - Gyeongsangbuk-do Province and Pohang City. G.S. acknowledges financial support from the Spanish MINECO grant PID2021-127726NB-I00, and PID2024-161474NB-I00 (MCIU/AEI/FEDER, UE), and from QUITEMAD-CM TEC-2024/COM-84. H.-C.Z. and G.S. also acknowledge support from the Grant IFT Centro de Excelencia Severo Ochoa CEX2020-001007-S funded by MCIN/AEI/10.13039/501100011033, and from the CSIC Research Platform on Quantum Technologies PTI-001 and the QUANTUM ENIA project Quantum Spain funded through the RTRP-Next Generation program under the framework of the Digital Spain 2026 Agenda. J.M.-V. thanks the financial support of Spanish MINECO grant PID2024-155685NB-C22 and Fundaci\'on S\'eneca de la Regi\'on de Murcia FSRM/10.13039/100007801 (22581/PI/24).





\bibliographystyle{JHEP}
\bibliography{main.bib}

\end{document}